\documentstyle[bo99,epsfig]{article}

\title{The Relativistic Astrophysics Explorer:\\ A New Mission for X-Ray Timing}

\author{P.~Kaaret$^{(1)}$, J.~Grindlay$^{(1)}$,
F.K.~Lamb$^{(2)}$, E.H.~Morgan$^{(3)}$, J.H.~Swank$^{(4)}$,
W.~Zhang$^{(4)}$}

\affil{1) Harvard-Smithsonian Center for Astrophysics\\ 
2) University of Illinois\\ 
3) Massachusetts Institute of Technology\\
4) NASA/Goddard Space Flight Center}

\begin{document}

\maketitle

\begin{abstract}

The great success of the Rossi X-Ray Timing Explorer (RXTE)
has given us a new probe to study strong gravitational fields
and to measure the physical properties of black holes and
neutron stars.  Here, we describe a ``next-generation'' x-ray
timing mission, the Relativistic Astrophysics Explorer (RAE),
designed to fit within the envelope of a ``medium-sized''
mission. The main instruments will be a narrow-field x-ray
detector array with an area of at least 60,000~cm$^2$ equal
to ten times that of RXTE, and a wide-field x-ray monitor
with good sensitivity and few arcminute position resolution. 
We describe the design of the instruments and the science
which will be possible with a factor of ten increase in
collecting area.

\end{abstract}

\section{Scientific Motivation for a New X-Ray Timing Mission}

The Rossi X-Ray Timing Explorer (RXTE) has made substantial
and unique contributions to the study of the behavior of
matter in strong gravitational fields near accreting compact
objects, the formation of relativistic jets, the emission
mechanisms of active galactic nuclei, the evolution of
neutron stars in binaries, the x-ray emission regions in
cataclysmic variables, and many other aspects of high-energy
astrophysics  (for a review see Bradt 1999).  The key feature
of RXTE is a large effective area x-ray detector coupled with
a high telemetry bandwidth.  The prowess of RXTE for fast
timing opened a new ``discovery space'' in rapid x-ray
variability, allowing timing studies at the dynamical time
scales of the innermost orbits around stellar mass compact
objects, and leading to the discovery of millisecond
quasiperiodic oscillations from accreting neutron stars and
black holes.  The large x-ray detector area also made
possible many other advances such as the discovery of
coherent millisecond pulsations from an accreting neutron
star, and the study of rapid spectral variations, such as the
$\sim 200 \rm \, s$ cycles in the microquasar GRS~1915+105
related to ejection of the inner regions of the accretion
disk.

The great success of the Rossi X-Ray Timing Explorer (RXTE)
is a strong indication that further progress in x-ray timing
will lead to new scientific advances. Here, we describe a
next generation x-ray timing mission which would offer an
order of magnitude increase in x-ray timing capabilities via
an x-ray detector with a geometric area of at least
60,000~cm$^2$, equal to ten times that of RXTE.  The most
important advances made with this order of magnitude increase
in collecting area are likely to be true discoveries and thus
cannot be anticipated.  However, an order of magnitude
increase in area would benefit many scientific
investigations.  Here, we describe three particular examples.

{\bf Fast quasiperiodic oscillations from black hole
candidates} (BHCs) have been discovered in three systems with
frequencies of 67-300~Hz (Remillard et al. 1999).  The fast
QPOs from BHCs are rather weak (rms amplitudes near 1\%) and
difficult to study in detail with RXTE.  A number of models of
the QPOs have been proposed, all of which involve strong-field
general relativistic effects, but distinguishing amongst the
various models will be difficult with the RXTE data.  The
increase in the photon statistics with RAE would make possible
much more accurate measurements of the QPO parameters and
their variations with time or correlations with spectral or
other timing parameters.  This may lead to a unique
identification of the QPO generation mechanism.  Understanding
these QPOs would provide a unique probe of strong-field
gravity.

{\bf Millisecond oscillations in x-ray bursts} have been
discovered from a number of neutron stars.  The oscillations
have periods in the range 1.7-3~ms and are interpreted as due
to inhomogeneous nuclear burning of matter initially located
on the neutron star surface.  The burst oscillations provide
a means to constrain the neutron star mass-radius relation. 
Currently, the best constraint comes from a deep modulation
($75\% \pm 17\%$) seen in the initial 62.5~ms of one burst
(Strohmayer et al. 1998).  RAE would detect roughly 1000
counts in each oscillation cycle near the peak of a typical
bright burst.  This would permit detailed examination of
individual oscillation cycles and allow accurate measurement
of the modulation amplitude in the first few oscillation
cycles.  Both our understanding of the burst oscillations and
constraints on the neutron star mass-radius relation would
improve.

{\bf Eclipse mapping} of the accreting magnetic white dwarf XY
Arietis showed that the x-ray flux emerges from eclipse egress
in $< 2$~s (Hellier 1997).  For the previous 15 years, the
fraction, $f$, of the white dwarf surface involved in x-ray
emission had been debated with values ranging from 0.001 to
0.3.  Hellier's result, obtained by combining 20 RXTE
observations, shows that $f < 0.002$.  Using RAE, an accurate
estimate could be made of the emitting region location on each
egress which would allow direct mapping of movement of the
emitting region.  Similar mapping can also be done in neutron
star and black hole binaries.  The best constraints currently
available on the size of the x-ray emitting regions in black
hole systems come from x-ray dips (e.g. Tomsick et al. 1997).
RAE would lead to significant advances in mapping x-ray
emission from many different x-ray sources.

\section{Mission Overview}

The Relativistic Astrophysics Explorer (RAE) will consist of
two scientific instruments: a large area x-ray detector and a
wide-field x-ray monitor.  RAE will be designed to have
telemetry sufficient to transmit the large event rate and
flexible operations with multiple repointings each day to
permit study of transient sources and rare states of known
sources.

The goal for the large area x-ray detector is to provide an
order of magnitude increase in x-ray timing capabilities
relative to RXTE.  The design goals are: a useful detector
area of at least 6~m$^2$, sensitivity from 2~keV to 30~keV,
absolute timing better than $10 \, \rm \mu s$, minimal dead
time effects for sources 10 times as bright as the Crab
nebula, an energy resolution of 1.2~keV (preferably 300~eV)
at 6~keV, no imaging, and a field of view of $1^{\circ}$ or
smaller.

All-sky x-ray monitoring is needed for several reasons. 
First, the x-ray monitor provides continual long-term light
curves.  As many x-ray sources are highly variable, knowledge
of the long term behavior in important in understanding the
physical nature of the sources and in placing pointed
observations in the context of the source state.  Second, an
x-ray monitor provides a means to trigger pointed observations
when a selected source reaches a state of particular
interest.  Finally, an x-ray monitor allows discovery of new
sources or new, unpredicted, outbursts of known sources.  Many
of the sources of interest are transients with unknown or
irregular recurrence intervals.  An x-ray monitor is essential
to detect transient events.  The design goal for the x-ray
monitor is a sensitivity of several mCrab for daily
observations, sufficient to monitor a large sample of AGNs
($\sim 40$) on a daily basis.

\section{Large Area X-Ray Detector Array}

An effective area of 6~m$^2$ will require a total geometric
detector area near 10~m$^2$.  A detector with a cross
sectional area of 10~m$^2$ and a thickness of 0.75~m$^2$ fits
within the 3~m diameter fairing of a two-stage Delta II.
Thus, a 6~m$^2$ detector can be accommodated in a
``medium-sized'' mission without a deployment mechanism.

The x-ray detector must have low mass per unit effective
area, reasonable cost, highly reliable and stable operation,
efficient rejection of particle backgrounds, and good energy
resolution.  After extensive review of the available detector
technologies, we have selected silicon detectors as the most
promising candidate for large format x-ray astronomy
detectors.  A 2~mm thick silicon detector provides 40\%
efficiency up to 30~keV at a mass of 0.5~gm/cm$^2$; this
compares favorably to PCA on XTE at a mass of 90~gm/cm$^2$. 
Silicon is widely used and can be obtained at low cost due to
large economies of scale; 10~m$^2$ of commercially available
silicon strip detectors (see below) can be  procured for less
than US\$4M. Silicon has a low ionization potential which
leads to good energy resolution and allows silicon detectors
to be operated without internal amplification.  While the
lack of internal amplification mandates the use of low
capacitance detectors and low noise electronics for good
performance, it also eliminates the need for high voltage and
facilitates reliable and stable operation.

Silicon detectors can be configured in different geometries
including PIN diodes, silicon strip detectors, and silicon
drift chambers.  Silicon strip detectors (SSDs) are widely
used in particle physics. SSDs offer one-dimensional imaging
which may provide a means of effective discrimination against
particle backgrounds.  However, SSDs employ charge collection
strips which run the length of a wafer and, thus, have
relatively high capacitance which leads to high electronic
noise and poor energy resolution.  Given mission constraints,
we estimate that the best energy resolution achievable with
SSDs will be 1--2~keV at 6~keV (see Costa et al.\ these
proceedings).

Silicon drift chambers (SDCs) (Gatti \& Rehak 1984) have an
internal electric field arranged so that electrons, produced
by interaction of radiation within the silicon, drift toward a
single charge collection point.  SDCs have a great advantage
in that the readout electrode can be made very small even if
the detection area is large; thus, the readout capacitance is
small, which leads to low readout noise and good energy
resolution.  Cylindrical SDCs, in which a single detector
electrode is used to collect charge from a cylindrical drift
region, are particularly well-optimized for x-ray spectroscopy
(Rehak et al.\ 1985; Lechner et al.\ 1996).  Excellent
performance has been demonstrated from  cylindrical SDCs in
the laboratory (Gauthier et al.\ 1994; Fiorini et al.\ 1997)
and in the field (Longoni et al.\ 1998).  Operating at
$-15^{\circ} \rm \, C$, a resolution of 155~eV (FWHM) at the
Mn-$K \alpha$ line has been obtained (Fiorini et al.\ 1997). 
The main question for x-ray astronomy is whether effective
particle background rejection can be achieved.  We currently
are engaged in a technology development program to study
application of SDCs to x-ray astronomy and to develop
effective techniques for charged particle background
rejection.

\section{Conclusion}

We have presented a conceptual design for a next generation
x-ray timing mission and identified key technologies which
must be developed.  The outstanding results from the Rossi
X-Ray Timing Explorer are strong motivation for a new mission
with an order of magnitude increase in x-ray timing
capabilities.


\end{document}